\documentclass[final]{elsart5p}
\usepackage{graphicx}
\usepackage{natbib}
\usepackage{amsmath}
\usepackage{amssymb}
\begin{document}

\begin{frontmatter}
\title{Theory of enhanced dynamical photo-thermal bi-stability effects in cuprous oxide/organic hybrid heterostructure}
\author{Oleksiy Roslyak},
\author{Joseph L. Birman}
\thanks{The project was supported in part by PCS-CUNY}
\address{Physics Department, The City College, CUNY\\
Convent Ave. at 138 St, New York, N.Y. 10031, 
USA}

\begin{abstract}
We theoretically demonstrate the formation of multiple bi-stability regions in the temperature pattern on the interface between a cuprous oxide quantum well and DCM2:CA:PS organic compound. The Frenkel molecular exciton of the DCM2 is brought into resonance with the $1S$ quadrupole Wannier-Mott exciton in the cuprous oxide by "solvatochromism" with CA. The resulting hybrid is thermalized with surrounding helium bath. This leads to strongly non-linear temperature dependence of the laser field detuning from the quadrupole exciton energy band which is  associated with the temperature induced red shift of the Wannier exciton energy. Numerical up and down-scan for the detuning reveals hysteresis-like temperature distribution. The obtained \emph{multiple} bi-stability regions are at least three orders of magnitude bigger ($meV$) than the experimentally observed bi-stability in bulk cuprous oxide ($\mu eV$). The effective absorption curve exhibits highly asymmetrical behavior for the Frenkel-like (above the $1S$ energy) and Wannier-like (below the $1S$ energy) branches of the hybrid.  
\end{abstract}
\end{frontmatter}
\section{Introduction}
The proposed of enhancement of nonlinear optical response from the quantum confined Wannier-Mott excitons (WE) by means of resonant hybridization with organic Frenkel excitons (FE) was first stated in the works of \cite{AGRANOVICH:1998,TIKHODEEV:2006,AGRANOVICH:2003}. The \emph{dipole-dipole} hybrid (DDH) is an appropriate coherent linear combination of large radius $a_B$ \footnote{small saturation density $\propto 1/a_B^2$ of the order of $10^{12} \; cm^{-2}$} \emph{dipole allowed} WE  and small radius but big oscillator strength $f^F$ \emph{dipole allowed} FE. The new hybrid is characterized by big oscillator strength and small saturation density (strong coupling regime). The expected hybrid optical nonlinearities are large because the ideal bosonic approximation starts to break down at the density of the hybrid close to the saturation density. Close to the saturation, due to overlapping between exciton wave functions, they exhibit rather strong exchange interaction and space filling factor. Compared to the bare WE nonlinear response, the hybrid response is enhanced roughly by the factor of the ratio of the FE and WE oscillator strength. 
\par
In our recent work \cite{ROSLYAK:2006} we demonstrated formation of a new type of the hybrid. It occurs between dipole forbidden but \emph{quadrupole allowed} $1S$ exciton in cuprous oxide quantum well and FE in an organic composite called DCM2:CA:PS. For the details on the organic see works of \cite{MADIGAN:2003} and \cite{BULOVIC:1999}. The FE formed on the DCM2 molecules is brought dynamically into resonance with the WE in the quantum well by means of "solid state solvation" [\cite{BALDO:2001}] mechanism. The \emph{quadrupole-dipole} hybridization (QDH) is of the same order of magnitude ($meV$) as that of the DDH: the small oscillator strength of the $1S$ quadrupole exciton [\cite{GROSS:1952}] is compensated by its strong spacial dispersion. This QDH exhibits the key features of big oscillator strength and narrow line width.
\par
In the present work we utilize the properties of the hybrid to enhance and modify nonlinear effects associated with the WM part of the hybrid. Due to the small radius of the $1S$ quadrupole $\approx 7$ $\AA$ the hybrid may achieve Bose condensation before it reaches the critical bulk density \footnote{we leave aside many questions about the ability for actual condensation of the hybrid for our future work}. Therefore, density dependent optical non-linearities which were important in the DDH case are negligibly small for the QDH. Hence, we focus our attention on another nonlinear optical phenomena generic to the cuprous oxide, namely, the photo-thermal bi-stability effect. 
\par
The photo-thermal bi-stability effect [\cite{ROSLYAK:2007,DASBACH:2004}] manifest itself for the density of the pumping laser as small as $P_{ex}=1 \; mWt$. Due to 'weak' interaction with acoustic phonons the energy of the WE will experience a red shift linear with the temperature [\cite{PASSLER:1998,VINA:1992,VARSHI:1967}]. The laser heating is balanced by the surrounding helium cooling. Hence, the resulting temperature on the interface is described by a nonlinear equation with more than one solution for some values of the laser detuning from the resonant energy level. This results in a hysteresis-like pattern for the temperature and absorption.        
\par
We have shown that the fact of hybridization with the organic enlarges the bi-stability region from $200 \; neV$ for the bulk cuprous oxide to the order of $meV$ for the hybrid due due to its large oscillator strength. Another remarkable effect of the hybridization is that the QDH exciton induces multiple highly asymmetrical photo-thermal bi-stability effects associated with two dispersion branches of the hybrid. The effective absorption maximum for the lower branch of the QDH exciton experiences a red shift of the order of $meV$ and strongly depends on the pumping intensity. 
\par
In the next section we propose a nonlinear absorption experiment and we give an appropriate mathematical model to observe the hybrid bi-stabilty. Our last section is devoted to discussion of the numerical results following from our model, and comparison with reported results for the bulk cuprous oxide [\cite{PASSLER:1998}].

\section{Nonlinear absorption by the hybrid quadrupole-dipole exciton}
 
\begin{figure}[htbp]
\centering
\includegraphics[width=8cm]{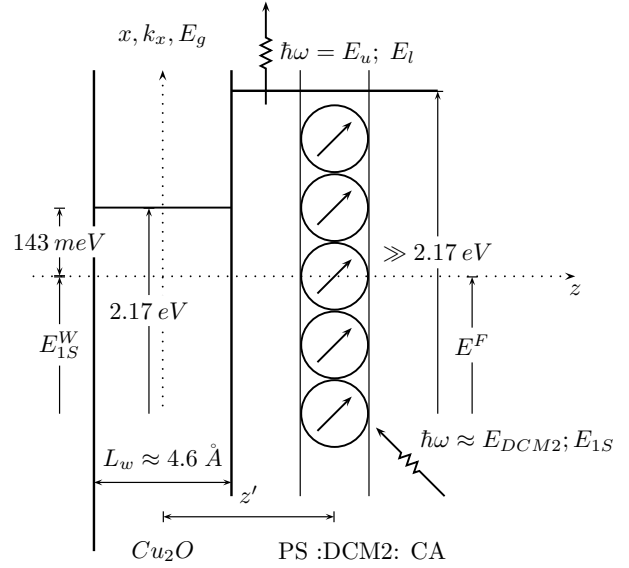}
	\caption{Schematic representation of a possible experimental set-up to nonlinear absorption by \emph{quadrupole-dipole} exciton. Here the inorganic $Cu_2O$ quantum well provides the $1S$ \emph{quadrupole} WE. The DCM2 part of the organic "solid state solute" provides \emph{dipole} allowed FE; the PS host prevents wave function overlapping between organic and inorganic excitons; CA under proper concentration allows tuning of the excitons into resonance.}
	\label{FIG:1}
\end{figure}

Our proposed configuration is shown in the following diagram (See Fig.\ref{FIG:1}). For the nonlinear absorption experiment let us consider a mono-layer of cuprous oxide placed upon a thin film of an organic compound We take the layer width $L_w$ approximately equal to the size of a unit cell $a=4.6 \; \AA$. The organic we propose is an 'solid state solvation' of the red laser dye DCM2 \footnote{[[2-methyl-6-2-(2,3,6,7-tetrahydro-1H, 5H - benzo[i,j] - quinolizin - 9 - yl) - ethenyl] - 4H - pyran - 4 - ylidene] propane dinitritle} embedded in a transparent host of polystyrene (PS) doped with the polar small molecule camorphic anhydride (CA) [\cite{BULOVIC:1999}]. One stationary laser (probe) is tuned around the $1S$ quadrupole transitions in the cuprous oxide $\hbar \omega \approx E_{1S}=2.05 \; eV$ (WE). 
\par
The pumping laser pulses of intensity $P_{ex}$ and energy $\hbar \omega = E_{DCM2}$ allow DCM2 molecular excitations. We consider the molecular excitation as a Frenkel exciton (FE). Therefore it can be viewed as a 2D lattice of dipoles $\mu_z^F = 20 \ D, \ f^F=0.45$ \cite{MADIGAN:2004} placed at $z'\approx L_w/2$. After the pumping pulse the FE experiences a red shift linear with the CA concentration due to non-resonant F\"{o}rster energy transitions to the CA molecules (See Fig.\ref{FIG:1}). 
\par
To achieve the red spectral shift of the FE into the resonance with the quantum confined WE, one has to adjust the CA concentration to $ \approx 22 \% $ [\cite{ROSLYAK:2006}]. To avoid complicated problems of the dynamics of the hybridizaton we assume that the FE and WE are in exact resonance once the DCM2 enrgy is in close proximity to the WE enrgy $E_{DCM2}-E_{1S} \leq \Gamma_{k}$. Now, we introduce the quadrupole-dipole coupling parameter [\cite{ROSLYAK:2006}]:
\begin{equation}
\label{EQ:2_1}
\Gamma_{k}=\frac{8\sqrt {2\pi } }{\left( 
{\varepsilon + \tilde {\varepsilon }} \right)L_w 
}\frac{ke^{-k{z}'}sinh\left( {\frac{L_w k}{2}} \right)}{\left( {1+\left( 
{\frac{kL_w }{2\pi }} \right)^2} \right)}\frac{Q_{xz} \mu_z^F }{a_F 
a_B L_W }
\end{equation}
Here $a_F$, $a_B$ are the Bohr radius of the FE and WE; $\tilde {\varepsilon }$ and $\varepsilon$ are the corresponding bulk dielectric constants of the organic and cuprous oxide. The quadrupole transition matrix element $Q_{xz}$ may be estimated from the corresponding bulk oscillator strength $f_{[110]}=3.6\times10^{-9}$ via the following identity [\cite{MOSKALENKO:2002}]:
\[
f_{xz,k_0 } =\frac{4\pi m_0 E_g }{3e^2\hbar ^2}\left( {\frac{a_B 
}{2\lambda a}} \right)^3{\rm {\bf e}}_z \cdot {\rm {\bf k}}_{0,x} \cdot Q_{x,z} 
\]
Here the energy gap of cuprous oxide is denoted as $E_g = 2.173 \; eV$ taken at $T=1.7 \ K$ of surrounding helium; $\lambda $ represents the effect of the quantum confinement, and $\left| {{\rm {\bf k}}_0 } \right|=2.62\cdot 10^5cm^{-1}$. The maximum value of the coupling parameter is about $2 \; meV$ and for the hybridization to be effective it must be of the same order of magnitude as the dissipative width of the hybrid $\gamma \hbar$.
\par
We assume that the cuprous oxide has purity of $99.99 \%$ with the reported line-width of $\hbar \gamma_{1S} = 0.1 \; meV$  i.e. pico-second lifetime [\cite{SHEN:1996}] while the FE life-time is of the order of $ns$. In this case the hybrid life-time is dominated by its inorganic part $\hbar \gamma \approx \hbar \gamma_{1S}$.
\par
The hybridization will strongly affect the absorption of the probe laser and consequently, it will also modify the photo-thermal bi-stability characteristics of the cuprous oxide. To study this effect, let us assume that the whole system is placed into helium bath with the temperature $T_{bath}=1.7 \; K$. The temperature of the exciton gas $T$ at the interface between the organic and inorganic is determined by the balance between the photo-thermal heating due to the exciton absorption and the cooling by the helium bath. 
\par
We will assume that the exciton gas is in equilibrium with the helium bath (thermalization) \footnote{To assure thermalization of the hybrid excitons we propose to pump the sample with consequent laser pulses coming one after another withing the hybrid life-time.}. For thermalized excitons one can treat the temperature $T$ as just another parameter of the system.
\par
The rate of helium cooling is given by the following linear term: $-C \left({T-T_{bath}}\right)$; the helium thermo-conductivity is denoted as: $C=2.8\times 10^{3}\  s^{-1} $ [\cite{DASBACH:2004}]. To find the heating term we derive below a semi-classical expression for the photo-thermal absorption of the hybrid. 
\par
The Hamiltonian of the system can be written in the rotating wave approximation \footnote{The rotating-wave approximation has been invoked by eliminating the anti resonant terms of the interaction with light; this approximation limits the calculation only to the resonant features of the response.} as following:
\begin{equation}
	\label{EQ:2_3}
	\begin{split}
	\tilde{H} = \delta b_\mathbf{k}^\dag  b_\mathbf{k}  + \Delta B_\mathbf{k}^\dag  B_\mathbf{k} + \Gamma _\mathbf{k} \left( {B_\mathbf{k}^\dag  b_\mathbf{k}  + B_\mathbf{k} b_\mathbf{k}^\dag  } \right) + \\
	+ \sqrt N \mu ^F \left( {E b_\mathbf{k}^\dag   + E b_\mathbf{k} } \right)
	\end{split}
\end{equation}
Here $ \delta = E_{DCM2} - \hbar \omega $ and $ \Delta = E_{1S} - \hbar \omega $ are the detunings of the laser frequency from the FE and WE respectively; $ N $ is the number of unit cells in the organic; $ E $ is the electric field of the incoming light; $B_\mathbf{k}$ and $b_\mathbf{k}$ are annihilation operators for the WE and FE respectively.
\par
\begin{figure}[htbp]
\centering
	\includegraphics[width=8cm]{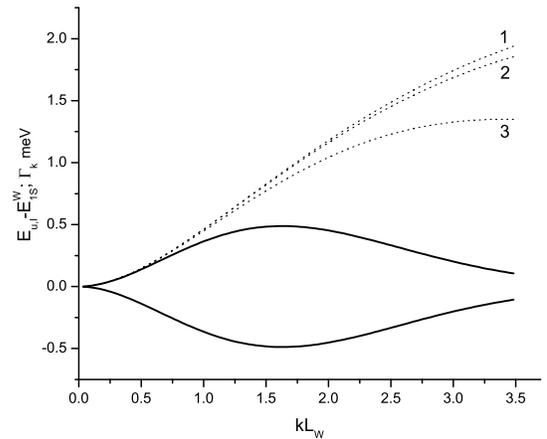}
	\caption{The solid lines represent upper and lower branches of the quadrupole-dipole hybrid dispersion when the coupling is calculated in the parabolic approximation and the induced Stark effect is taken into account; the dash lines correspond to the coupling parameter for different approximations: (1) - infinite IQW and the induced Stark effect is neglected, (2) - parabolic approximation, (3) - infinite IQW with the induced Stark effect treated as perturbation}
	\label{FIG:3}
\end{figure}

Because the frequency of the laser pulses is much smaller than $\omega$, one can consider the laser field turned on adiabatically. Then, the equations of motion for any operator can be solved in a very straightforward manner. Apply to that operator the unitary transformation that diagonalizes (\ref{EQ:2_3}) and then introduce an imaginary part to all detunings, positive for creation operators and negative for annihilation operators. Since the Hamiltonian involves only Bose operators and is quadratic, the unitary transformations that diagonalizes it can be obtained in an analytic form. The diagonalization procedure can be written as:
\[
U^{ - 1} \tilde HU = e^{ - S_2 } \left( {e^{ - S_1 } \tilde He^{S_1 } } \right)e^{S_2 } 
\]
where 
$
S_1  = \left( {x_{\bf{k}} b_{\bf{k}}^\dag   - x_{\bf{k}}^* b_{\bf{k}} } \right) + \left( {y_{\bf{k}} B_{\bf{k}}^\dag   - y_{\bf{k}}^* B_{\bf{k}} } \right)
$
eliminates the linear terms in 
$
\left( {b_{\bf{k}}^\dag   + b_{\bf{k}} } \right)
$. 
The transformation coefficients age given by
\[
x_{\bf{k}}  = \sqrt N \frac{{\left( {\mu ^F E} \right)\Delta }}{{\Delta \delta  - \Gamma _{\bf{k}}^2 }}, \  y_{\bf{k}}  = \sqrt N \frac{{\left( {\mu ^F E} \right)\Gamma _{\bf{k}} }}{{\Delta \delta  - \Gamma _{\bf{k}}^2 }}
\]
The bilinear cross term 
$
{B_{\bf{k}}^\dag  b_{\bf{k}}  + B_{\bf{k}} b_{\bf{k}}^\dag  }
$
can be diagonalized away through a rotation by $\phi _{\bf{k}}$ in the FE-WE coordinate space [\cite{MUKAMEL:1995}] of the form:
\begin{equation*}
S_2  = \phi _{\bf{k}} B_{\bf{k}} b_{\bf{k}}^\dag   - \phi _{\bf{k}} B_{\bf{k}}^\dag  b_{\bf{k}}, \
\tan \left( {2\phi _{\bf{k}} } \right) = \frac{{2\Gamma _{\bf{k}} }}{{\delta  - \Delta }} 
\end{equation*}
The overall diagonalized Hamiltonian has the form:
\begin{equation}
\label{EQ:2_4}
\begin{split}
\left\{ {\frac{{\delta  + \Delta }}{2} + \sqrt {\left( {\frac{{\delta  - \Delta }}{2}} \right)^2  + 4\Gamma _{\bf{k}}^2 } } \right\}B_{\bf{k}}^\dag  B_{\bf{k}} + \\
+ \left\{ {\frac{{\delta  + \Delta }}{2} - \sqrt {\left( {\frac{{\delta  - \Delta }}{2}} \right)^2  + 4\Gamma _{\bf{k}}^2 } } \right\}b_{\bf{k}}^\dag  b_{\bf{k}} 
\end{split}
\end{equation}
The factors in front of number operators $B_{\bf{k}}^\dag  B_{\bf{k}}$ ($b_{\bf{k}}^\dag  b_{\bf{k}}$) represent the upper branch and lower branch of the hybrid at exact resonance between FE, WE and the photon $E_{1S}=E_{DCM2}=\hbar\omega$. The dispersion is illustrated on Fig.\ref{FIG:3} [\cite{ROSLYAK:2006}].   
\par
With the use of this unitary transformation, the direct contribution of the exciton transition to the induced polarization per unit area of the interface is
\begin{equation}
\label{EQ:2_5}
P = \left\langle i \right|U^{ - 1} \left( {\frac{{\sqrt N }}{S}\mu ^F \left( {b_{\bf{k}}  + b_{\bf{k}}^\dag  } \right)} \right)U\left| i \right\rangle 
= \frac{N}{S}\frac{{\left( {\mu ^F } \right)^2 \Delta }}{{\Delta \delta  - \Gamma _{\bf{k}}^2 }}E
\end{equation}
Here the initial state $ \left| i \right\rangle $ corresponds to the FE exciton, as we neglected the quadrupole-light interaction. The hybrid life-time is introduced into the expression above through the complex part of the detunings    
$
\delta  \to \delta  - i\hbar \gamma, \  \Delta  \to \Delta  - i\hbar \gamma 
$ as discussed above.
\par
Also, due to the localization of the Frenkel non-resonant excitation, we take the FE into account as corrections to the local field. Assuming the organic to be an isotropic medium, one can write total electric field to be
\[
E = E_{laser}  + \frac{{4\pi }}{3}\left( {\alpha _{res}  + \alpha _{non - res} } \right)E_{laser} 
\]
The second term represents the local field contribution, where $ \alpha_{non - res} $ is the background polarizability (defined by non-resonant excitations in the organic) , while $ \alpha_{res} $ is the direct exciton contribution to the polarizability corresponding to ( \ref{EQ:2_5} ). The last equation can be rewritten as:  
\[
E = \frac{{\left( {\tilde \varepsilon  + 2} \right)/3}}{{1 - \left( {4\pi \alpha _{res} /3} \right)\left( {\tilde \varepsilon  + 2} \right)/3}}E_{laser}
\]
where the Lorentz-Lorenz relationship
\[
\frac{{4\pi }}{3}\alpha _{non - res}  = \frac{{\tilde \varepsilon  - 1}}{{\tilde \varepsilon  + 2}}
\]
has been used. Thus, the local field corrections results in the exciton frequency shift 
$
\delta  \to \delta  - \frac{{4\pi }}{3}\frac{N}{S}\left( {\mu ^F } \right)^2 \left[ {\frac{{\tilde \varepsilon  + 2}}{3}} \right]
$. The susceptibility is given by: 
\begin{equation*}
\chi  = 
\frac{N}{S}\frac{{\left( {\mu ^F } \right)^2 \Delta }}{{\Delta \delta  - \Gamma _{\bf{k}}^2 }}
\end{equation*}
Taking into account that $ N/S \approx 1/a_F^2 $, the absorption coefficient is given by:
	\begin{equation}
	\label{EQ:2_6}
	\Im \chi =
	\frac{f^F e^2 \gamma  \Delta ^2 \hbar }{2 c a_F^2
   \left(\left(\delta  \Delta -\Gamma_{\bf{k}}
   ^2\right)^2+\gamma ^2 \Delta ^2 \hbar ^2\right)
   \sqrt{\tilde{\varepsilon }} m_0}
  \end{equation}
Let us now make an important observation, that this absorption coefficient is fundamentally different from that for the case of the DDH. In case of the QDH, one has $\hbar \gamma_F \gg \hbar \gamma_{hyb} \gg \hbar \gamma_1S$ while for a dipole-dipole hybridization they are of the same order of magnitude. So, when $\Delta \to 0$ the absorption for the QDH is zero even for a smallest coupling.
\par
For the QDH using the given absorption coefficient (\ref{EQ:2_6}), the transmission through the sample of an arbitrary thickness (See Fig.\ref{FIG:1}) can be written as:
\begin{equation}
\label{EQ:2_8}
Tr = 1 - \exp \left( { - \frac{{\left( {\gamma \hbar \Delta} \right)^2 }}{{\left( {\delta \Delta  - \Gamma _{\bf{k}}^2 } \right)^2  + \left( {\gamma \hbar \Delta} \right)^2 }}\frac{z'}{l}} \right)
\end{equation}
Here $ z' $ can be viewed as the width of the narrow strip around the interface where the hybridization occurs and $ l = \frac{\gamma \hbar c \sqrt{\epsilon_{\infty}}}{f^F E_{1S}^2} $ is the hybrid absorption length.
\par 
Due to the big oscillator strength and narrow line width of the hybrid the absorption length $l$ is extremely small. This is balanced out by the fact that the hybridization occurs only in a narrow strip around the interface between the organic and inorganic. The rest of the crystal is approximately transparent. 
So we can assume that the absorbing region is equal to the Bohr radius of the hybrid, i.e. $z' = a_B =7 \; \AA$, therefore $z'/l \approx 1 $. The value of the wave vector ${\bf k}$ is controlled by the oblique angle of incidence for the probe laser.
\par
The "weak" interaction of the WE in the quantum well with the LA phonons leads to a red shift of the semiconductor energy gap and therefore to the red shift of the WE band gap [\cite{PASSLER:1998,VINA:1992,VARSHI:1967}] and does not affect the binding energy of the $1S$ exciton [\cite{SNOKE:1990}]. Hence, for small deviation of the exciton gas temperature from the helium bath temperature one can consider the shift to be linear [\cite{ROSLYAK:2007,DASBACH:2004}]:
\begin{equation}
\label{EQ:2_9} 
E^W \left( {T} \right) = E_{1S}\left(0\right)-\kappa\left({T-T_{bath}}\right)
\end{equation}
where $\kappa=0.3 \; \frac{\mu eV}{K}$.
\par
The final temperature change given by the laser heating and helium cooling can be written as:
\begin{equation}
\label{EQ:2_10}
\frac{{dT}}{{d\tau }} = HP_{ex} \left[ {1 - {\rm Tr}\left( \tau, T  \right)} \right] - C\left[ {T - T_{bath} } \right]
\end{equation}
Here the phenomenologically introduced constants $ H=5.0\times 10^6 \ K s^{-1} W^{-1} $ and $ C=2.8\times 10^{3}\  s^{-1} $ giving the heating and cooling rate, respectively [\cite{DASBACH:2004}]. The transmission (\ref{EQ:2_8}) depends on the temperature through the detuning $\Delta\left( T \right)$ according to the equation (\ref{EQ:2_9}). The equilibrium temperature is given by the stationary solutions of the equation (\ref{EQ:2_10}). This will govern the temperature and absorption bi-stability.

\section{Results and discussion}

The QDH structure offers a more complicated picture of the nonlinear photo-thermal effect then just a slab of cuprous oxide itself. As we now show it is not a mere enhancement of the bi-stability effect but the appearance of two or more well pronounced bi-stability regions in temperature and transmission (absorption) spectra. In case of the QDH the photo-thermal rise of the temperature leads not only to the red shift of the $1S$ exciton energy band but to an effective breaking of the resonance with the FE and reduction of the hybridization.
\par
\begin{figure}[htbp]
\centering
	\includegraphics[width=8cm]{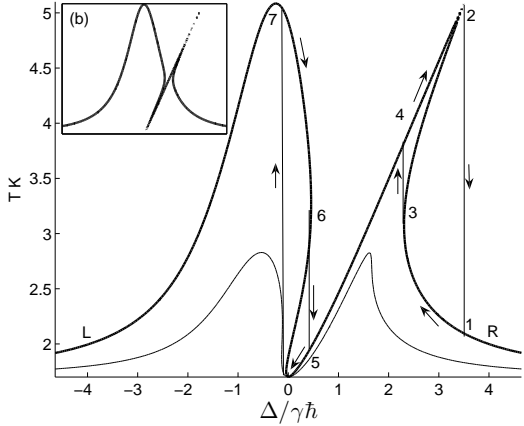}
	\caption{The temperature of the sample in K, versus the detuning in units of the hybrid exciton line-width $\gamma \hbar$. The coupling parameter is taken to be equal to the line width of the hybrid and the subplot (b) corresponds to the Stark reduced coupling. The thick solid curve corresponds to the laser intensity $P_{ex}=3 \; mWt$ and maximum temperature $5 \; K$, and another thin curve stands for the intensity $P_{ex}=1\; mWt$ and maximum temperature $2.5 \; K$}
	\label{FIG:4}
\end{figure}

Many properties of the temperature, absorption and effective dispersion of the hybrid system can be obtained even before a numerical solution for the system of equations (\ref{EQ:2_9},\ref{EQ:2_10}). The right hand side of the equation (\ref{EQ:2_10}) can be visualized as an implicit function of the temperature and detuning (See Fig.\ref{FIG:4}). 
The actual temperature of the system depends on the way one changes the frequency of the probe laser. If one lowers the frequency starting from the detuning $2 \; meV$ (point L on Fig.\ref{FIG:4}) off resonance, then the system follows the path $R \to 1 \to 3 \to 4 \to 5 \to 0 \to 7 \to L$ to assure minimum temperature change with the changing detuning. If one increases the frequency  starting from the detuning $-2 \; meV$ (point R on the Fig.\ref{FIG:4}), then the system follows the path $L \to 7 \to 6 \to 5 \to 4 \to 2 \to 1 \to R$.
\par
\begin{figure}[htbp]
\centering
	\includegraphics[width=8cm]{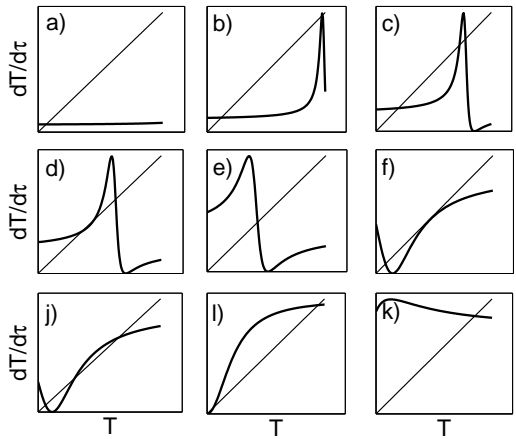}
	\caption{The numerical solution of the nonlinear equation (\ref{EQ:2_10}) is given crossing of the cooling (straight line starting from the origin) and the detuning dependent photo-heating. The figure a)-k) correspond to the different laser detuning from the low temperature limit of the $1S$ quadrupole exciton as $\frac{\Delta}{\hbar \gamma}=5;\;3.7;\;3;\;2.5;\;1.8;\;0.5;\;0.4;\;0;\;-0.8$ in units of the damping parameter.}
	\label{FIG:5}
\end{figure}

On the other hand one can consider bi-stability as a possibility for the the system (\ref{EQ:2_9},\ref{EQ:2_10}) to have more then one stationary solution. This non-linearity grows with $P_{ex}$. To illustrate the concept it is convenient to plot the heating and cooling rate as a function of the temperature. Then the stationary temperature is given as the intercept points ( See Fig.\ref{FIG:5}).
\par
\begin{figure}[htbp]
\centering
	\includegraphics[width=8cm]{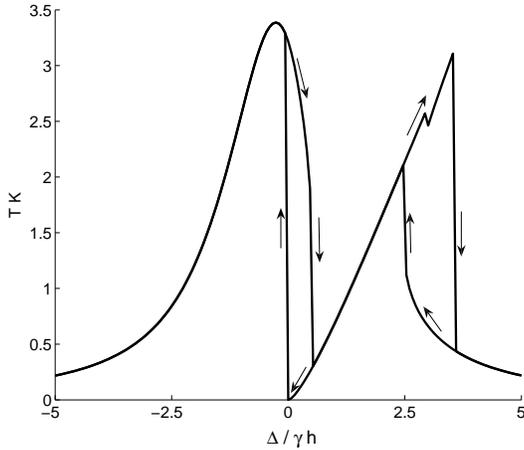}
	\caption{The numerical results for the temperature of the sample in K, versus the detuning in meV, $P_{ex}=3 \; mWt$. The coupling parameter is taken to be equal to the line width of the hybrid. The bi-stability is a result of multiple solutions and reveal itself as a sudden drop or raise in the temperature depending on the direction of the scan.}
	\label{FIG:6}
\end{figure}

Now let us focus on the different branches of the hybrid. It is clear from the Fig.\ref{FIG:4} and Fig.\ref{FIG:6} that the response from two branches is asymmetrical. Further on we are going to refer to the lower branch as Wannier-like and to another  as Frenkel-like. Both of them are defined by the interplay between the thermo-induced red shift of the WE exciton $\kappa \left({T-T_{bath}}\right)$ and the excitation detuning $\Delta$. 
\par
Indeed, the maximum of the absorption given by the poles of the expression (\ref{EQ:2_6}):
\begin{equation}
\label{EQ:3_1}
2\delta_{\pm} = \kappa \left({T-T_{bath}}\right) \pm \sqrt{ \kappa^2 \left({T-T_{bath}}\right)^2+4\Gamma_\mathbf{k}^2}
\end{equation} 
If the coupling between WE and FE is dominant: $\kappa^2 \left({T-T_{bath}}\right)^2 \ll 4\Gamma_k^2$, then both branches are symmetrical and the maximum of absorption corresponds to the usual hybrid dispersion $\delta_{\pm}=\pm \Gamma_k$. Otherwise,  strong asymmetry reveals itself. The case $\delta_+ \to \kappa \left({T-T_{bath}}\right)$ defines the maximum of Wannier-like bi-stability, dominated by the temperature induced red shift. The Frenkel-like bi-stability corresponds to $\delta_- \to 0$ and is dominated by the hybridization effect. 
\par
For the up-scanned Wannier-like branch when the excitation energy $\delta$ catches up with the escaping exciton resonance $\delta_+$ the actual temperature cut-off (Path $2 \to 1$ on the Fig.\ref{FIG:4} and Fig.\ref{FIG:5},(l))  occurs when the laser heating reaches its maximum $\delta_+=\Delta$. When the excitation energy increases further the absorption reduces and the sample cools rapidly. On the down-scan the sample is cold even beyond the cut-off energy (the temperature is close to the lower cross-point on  Fig.\ref{FIG:5},(l)). 
\par
In the same fashion the temperature rises rapidly on the Frenkel-like branch (Path $0 \to 7$ on the Fig.\ref{FIG:4} and Fig.\ref{FIG:5},(b))with a down-scan. The temperature goes down and the detuning between FE and WE reaches its minimum and an abrupt heating occurs. For the up-scan the sample is heated enough (close to the upper cross-point on  Fig.\ref{FIG:5},(b)) to maintain large detuning between the FE and WE but drops down rapidly when the FE and WE detuning reaches its maximum $\delta_-=\Delta$ (See Fig.\ref{FIG:5},(d)).
\par
The bi-stability effect and the temperature rise of the hybrid is greatly enhanced ($meV$) compared to bulk cuprous oxide itself ($\mu eV$) mainly due to the fact of gaining oscillator strength from the organic part of the hybrid.
\par
\begin{figure*}[htbp]
\centering
	\includegraphics[width=10cm]{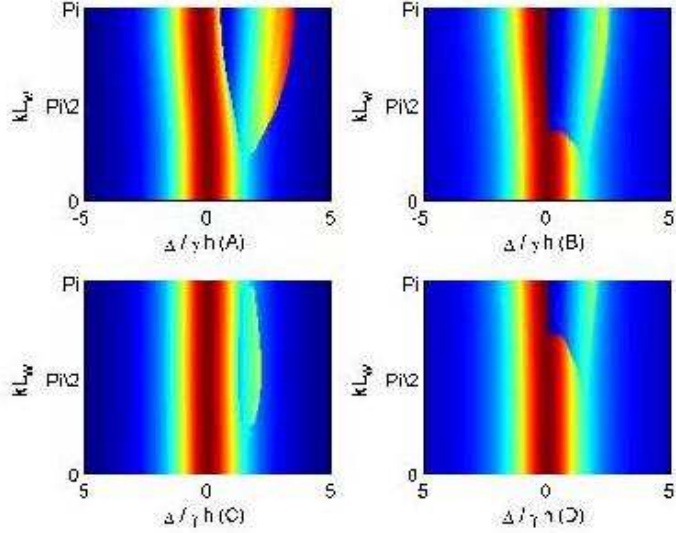}
	\caption{The temperature of the interface between organic DCM2:CA:PS and inorganic cuprous oxide QW for different detuning of the laser field from the energy of the $1S$ quadrupole exciton take at $T=T_{bath}$ and different wave vectors $k$ of the hybrid. The blue color corresponds to the temperature of the surrounding helium $1.7K$. The graphs represents the up (A) and down (B) laser scanning. The stark effect is taken into account on the up-scan (C) and down-scan (D) graphs.}
	\label{FIG:7}
\end{figure*}

Numerical search for a stationary solution of equation (\ref{EQ:2_10}) reveals a fine structure of the multiple bi-stability (See Fig.\ref{FIG:6}, Fig.\ref{FIG:7}). For the up-scan regime the distance between the solution branches $4 \to 2$ and $3 \to 2$ on the Fig. \ref{FIG:4} can become smaller then $\kappa \left({T-T_{bath}}\right)$ and the temperature drops down on the branch $3 \to 2$. For the same reason when the intensity is big enough and the coupling between the FE and WE is reduced by the Stark effect fine structure can appear on the branches $5 \to 0$ and $6 \to 0$ on the down-scan.
\par
The model described above is valid for small temperature raise from the $T_{bath}$ only. But it can be generalized for bigger temperature change by using a more elaborated non-linear temperature dependence of the WE exciton [\cite{VARSHI:1967}] and possible effect of the optical phonons [\cite{PASSLER:1998}]. Although, when one increases $P_{ex}$ to the point when the density of the hybrid is close to the saturation density then other nonlinear effects due to exciton-exciton scattering and possible condensation start to play a considerable role. Such non-linear effects are presently under investigation. 

\end{document}